\documentclass[10pt,aps,prb,twocolumn,amsmath,amssymb,superscriptaddress,longbibliography,floatfix]{revtex4-2}

\usepackage{amsmath}  
\usepackage{amssymb}  
\usepackage{graphicx}  
\usepackage{hyperref}   % Load hyperref LAST  
\usepackage{cleveref}   % Then cleveref (if used)  
\usepackage[utf8]{inputenc}
\usepackage[utf8]{inputenc}
\usepackage{placeins}
\FloatBarrier   % insert just before/after the problem area
\DeclareUnicodeCharacter{2212}{-}  % Fixes the minus sign issue
\DeclareUnicodeCharacter{2009}{\,} % Fixes thin space issue
\hypersetup{
  colorlinks=true,
  citecolor=Violet,
  linkcolor=Red,
  urlcolor=Magenta}
\usepackage{natbib,hyperref}
\usepackage{amsmath,epsfig,color,amssymb}
\usepackage{bbold}
\usepackage[dvipsnames]{xcolor}
\usepackage[export]{adjustbox}
\usepackage{comment}
\usepackage[titletoc,title]{appendix}
\newcommand{\beg}{\begin{equation}}
\newcommand{\en}{\end{equation}}

\newcommand{\sign}{\text{sign}}

\newcommand \bel  {\begin{align}}
\newcommand \enl  {\end{align}}

\newcommand{\braket}[2]{\left< #1 \vphantom{#2} \right|
  \left. #2 \vphantom{#1} \right>} % for Dirac brackets
\definecolor{new}{rgb}{.08,.05,.8}

\begin{document}

%\title{Effects of Strain-Induced Pseudogauge Fields on Exciton Dispersion, Transport, and Interactions in Transition Metal Dichalcogenides
%}
\title{Effects of Strain-Induced Pseudogauge Fields on Exciton Dispersion, Transport, and Interactions in Transition Metal Dichalcogenides Nanoribbons}

\author{Shiva Heidari}
\affiliation{Physics Department, City College of the City University of New York, NY 10031, U.S.A.}

\author{Shervin Parsi}
\affiliation{Physics Department, City College of the City University of New York, NY 10031, U.S.A.}
\affiliation{Physics Program, Graduate Center of City University of New York, NY 10031, U.S.A.}

\author{Pouyan Ghaemi}
\affiliation{Physics Department, City College of the City University of New York, NY 10031, U.S.A.}
\affiliation{Physics Program, Graduate Center of City University of New York, NY 10031, U.S.A.}

\begin{abstract}
We study the effects of strain on exciton dynamics in transition metal dichalcogenide (TMD) nanoribbons. Using the Bethe-Salpeter formalism, we derive the exciton dispersion relation in strained TMDs and demonstrate that strain-induced pseudo-gauge fields significantly influence exciton transport and interactions. Our results show that low-energy excitons occur at finite center-of-mass momentum, leading to modified diffusion properties. Furthermore, the exciton dipole moment depends on center-of-mass momentum, which enhances exciton-exciton interactions. These findings highlight the potential of strain engineering as a powerful tool for controlling exciton transport and interactions in nanoribbon-based TMD optoelectronic and quantum devices.
\end{abstract}

\maketitle

\section{introduction}
The realization of excitons with large binding energies in two-dimensional transition metal dichalcogenides (TMDs) present new opportunities for studying and utilizing exciton transport, even at room temperature \cite{lackner2021tunable,darlington2020imaging,rosati2021dark,wallauer2021momentum,huang2019robust,PhysRevB.97.245411}. A key challenge in this context is controlling exciton motion in a specific direction and over large distances, particularly at room temperature.

Another advantage of two-dimensional materials is the possibility of straining them in a controlled manner. The effects of strain on excitonic states and their diffusion have been studied both theoretically and experimentally \cite{dirnberger2021quasi, rosati2021dark, PhysRevLett.120.207401, kumar2024strain, rosati2020strain}. Most previous studies have explored the influence of strain on excitons in terms of local potentials that is, the shift in electronic energy bands \cite{dirnberger2021quasi,krustok2017local,PhysRevB.87.155304}. Strain also induces symmetry breaking and fine-structure modifications in excitons, as observed in recent studies \cite{PhysRevB.106.125303}.
Another distinct effect of strain in two-dimensional materials, extensively studied in graphene, is the generation of a pseudomagnetic field, which leads to the formation of flat Landau levels\cite{guinea2010energy,levy2010strain,nigge2019room}. This effect in graphene would lead to phenomena such as the fractional quantum Hall effect \cite{PhysRevLett.108.266801,PhysRevB.87.155422} and Landau-level excitons \cite{berman2022strain}. Despite their gapped band structure, TMDs can exhibit similar effects, where the conduction and valence bands transform into a set of Landau-like bands \cite{PhysRevB.92.195402,PhysRevB.86.241401,shayeganfar2022strain}.
Similar to graphene, this strain-induced effect in TMDs can be described in terms of pseudo-gauge fields. The strain-induced bands are separated by a finite gap, which is controlled by the magnitude of the applied strain. A few percent of lattice deformation can induce a gap on the order of room temperature \cite{lee2017strain}.

In this paper, we examine the effect of strain-induced modifications in the electronic band structure, resulting from pseudo-gauge fields, on excitonic transport in TMD nanoribbon. Our results indicate that such modifications in the electronic band structure lead to enhanced anisotropic exciton diffusion, well beyond the effects of strain-induced local potentials.
Another key property of electronic bands in some of TMDs is the presence of band inversion near the bottom of the bands in each valley \cite{krishnamurthi20211d, PhysRevB.93.125109,ugeda2018observation}. Although the presence of an even number of valleys renders the full band structure topologically trivial, the possibility of optically selective coupling with the bands in each valley leads to observable effects of nontrivial band topology within an individual valley \cite{PhysRevB.89.115413,maisel2023single}. We show that in the presence of band inversion, strain shifts the exciton energy band minimum to a finite momentum and stabilizes dark excitons, which naturally have a longer lifetime \cite{PhysRevResearch.3.043198}. These effects collectively support longer rang unidirectional exciton diffusion.

Beyond its influence on exciton transport, we show that strain can enhance exciton-exciton interactions. Under the type of strain discussed in this paper, the electronic states resemble the Hall electronic states in the Landau gauge \cite{Hbook}. Due to translation invariance along the nanoribbon, these states are labeled by their momentum in this direction and the maximum of their spatial wavefunction in the perpendicular direction is controlled by the momentum along the ribbon. As a result, excitons with finite center-of-mass momentum are formed by binding electrons and holes that are spatially separated in a plane, in direction prependicular to the nanoribbon. This structure, which leads to a large in-plane dipole moment, facilitates strong exciton-exciton interactions \cite{sun2022excitonic}.
Our results show that pseudo-gauge fields generated by specific strain patterns provide an efficient avenue for simultaneously facilitating exciton transport and strong exciton-exciton interactions. Consequently, strain engineering emerges as a powerful tool for realizing and utilizing novel excitonic phases in TMDs, through effects that go beyond the mere generation of local potentials.

The remainder of this paper is organized as follows. In Section II, we describe the electronic band structure in strained TMD nanoribbons. Section III presents our results on exciton dispersion which is derived using the Bethe-Salpeter formalism in strained TMDs and analyze the effects of strain-induced pseudo-gauge fields on exciton dispersion. In section IV we study the effect of the strain induced pseudo gauge fields on exciton transport in strained TMD nanoribbons and demonstrate how strain enhances unidirectional diffusion. Section V discusses the effect of strain on enhancement of exciton dipole moment which leads to stronger interaction between excitons. In Section VI, we summarize our key findings and discuss potential applications of strain-engineered TMD nanoribbons.

\section{Electronic band structure of strained TMD nano-ribbon}
To model the effects of strain on excitonic properties, we consider a monolayer transition metal dichalcogenide (TMD) nanoribbon with zigzag edges and finite width along the \( y \)-direction and open boundary along \( x \). The strain leads to a displacement of the lattice sites from the original position $\textbf{r}_0$ to $\textbf{r}_0+\textbf{r}_0.\nabla \textbf{u}$ \cite{PhysRevB.87.165131}.  The dominant effect of the strain on the electronic band structure results from the modifications in the hopping amplitudes between lattice sites. The accompanying change in the geometry of the lattice is commonly ignored since its effects are subdominant \cite{PhysRevB.92.195402,guinea2010energy}. An arc-shaped strain pattern corresponds to the displacement field \( (u_x, u_y) = (xy/R, -x^2/2R) \), where \( R \) is the radius of curvature. The effect of this type of strain shows up in the effective continuum Hamiltonian as a pseudo-gauge field which only depends only on $y$ and preserves the translation invariance along $x$ \cite{PhysRevB.92.195402}. In contrast to longitudinal or transverse uniaxial strain, which primarily result in local potential shifts, the arc-shaped profile leads to pseudomagnetic effects that significantly alter the band structure. A schematic of the system geometry, coordinate axes, and basis functions is shown in Fig.~\ref{fig1}. The effective Hamitlonain $H_0$ presented in Eq. \ref{H0} is then translationally invariant along the longitudinal \( x \)-direction, hence momentum along \( x \) is conserved and the system has a quasi-one-dimensional geometry. As shown in Eq. \ref{H0} the Hamiltonian is a function of momentum along $x$ direction (i.e., $k_x$) and position along $y$ direction.

Excitonic states arise from the attractive Coulomb interaction between electrons in the conduction band and holes in the valence band \cite{srivastava2015valley,RevModPhys.90.021001,PhysRevB.110.035425,PhysRevLett.113.076802}. In order to derive the exiton dispersion in strained TMDS we first present their electronic band structure. We use the effective continuum Hamiltonian $H_0$ for parallel momentum close to the center of each valley which is modified due to the strained induced modification of hopping parameter in the  Slater-Koster tight-binding model  \cite{PhysRevB.88.075409,PhysRevB.92.195402}. The effect of later modifications in the continuum Hamiltonian are presented in terms of pseudoguage fields and reads as:
    \begin{figure}[!htbp]
		\centering
		\includegraphics[width=\linewidth]{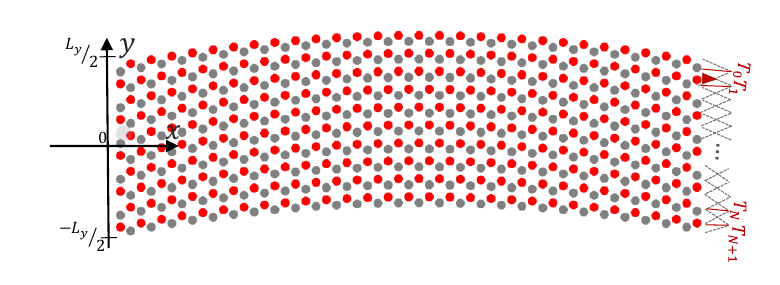} \caption{Schematic illustration of a strained monolayer TMD nanoribbon. The nanoribbon is finite along the \textbf{y}-direction (width \( L_y \)) and translationally invariant along the \textbf{x}-direction (length). The applied strain bends the ribbon into an arc shape described by the displacement field \( (u_x, u_y) = (xy/R, -x^2/2R) \), where \( R \) is the radius of curvature. This deformation generates a position-dependent pseudogauge field that modifies the electronic band structure.
        The trigonal basis function \( T_n(y) \) is defined as \( T_n(y) = L_y - |y - y_n|(N + 1) \), where \( |y| < L_y/2 \), and \( y_n = L_y\left(\frac{n}{N+1} - \frac{1}{2}\right) \) is the position of the \( n \)-th basis function. \( N \) is the number of basis functions, and \( L_y \) is the width of the nanoribbon along the y-direction. This function is used to satisfy the hard boundary conditions for an arc-shaped monolayer Transition Metal Dichalcogenides (TMDs) system. We note that \( \langle T_n | T_{n+1} \rangle \ne 0 \). 
    }
		\label{fig1}
	\end{figure}

\begin{equation} \label{H0}
\resizebox{1.0\hsize}{!}{$
  H_0=\begin{pmatrix}  
        {\cal V}_+(y)-\frac{\hbar^2}{4 m_0} (\alpha+\beta) \partial^2y & t_0 a_0 (k_x+\frac{e}{\hbar}A_{1,x})-t_0 a_0 \partial_y \\
        t_0 a_0 (k_x+\frac{e}{\hbar}A_{1,x})-t_0 a_0 \partial_y & {\cal V}_-(y)-\frac{\hbar^2}{4 m_0} (\alpha-\beta) \partial^2y
      \end{pmatrix}$}.
\end{equation}
 
The parameters in $H_0$ are $t_0=2.34$ eV corresponding to nearest neighbour hopping amplitude and $\alpha=-0.01$ and $\beta=-1.54$ are dimensionless parameteres which result from expansion of Slather-Kosher tight-binding Hamiltonian for parallel momentum $k_x$ close to the corners of BZ. $\hbar$ and $e$ are the reduced Planck constant and elementary charge, respectively and $a_0$ is the MoS$_2$ lattice parameter. The potential terms ${\cal V}_\pm(y)$ incorporates spin-orbit interaction which has a dominant diagonal form $S_zL_z$\cite{Roldán_2014}, leading to the possibility of treating the two values of $S_z$ separately, and is given by: \cite{PhysRevB.92.195402}
\begin{equation}
\begin{split}
    {\cal V}_\pm(y)&= \frac{(\Delta_0+\sigma \lambda_0)\pm (\Delta+\sigma \lambda)}{2}+{\cal D}_\pm(y)+\sigma \delta \lambda_\pm(y)  \\ &+\frac{\hbar^2}{4 m_0} \alpha (k_x+\frac{e}{\hbar}A_{2,x})^2+\frac{\hbar^2}{4 m_0} \alpha (k_x+\frac{e}{\hbar}A_{3,x})^2,
\end{split}
\end{equation}
where $\sigma=\pm 1$ is the spin index. The potential ${\cal V}_\pm(y)$ includes spin-dependent energy gaps ($\Delta_0$, $\Delta$)=$(-0.11,1.82)$ eV and spin-orbit parameters ($\lambda_0$, $\lambda$)=($69,1.82$) meV. The Spatially varying terms ${\cal D}_{\pm}(y)$ and $\delta\lambda_{\pm}(y)$ are given by \cite{PhysRevB.92.195402}, 
\begin{align}  \label{eqs}
D_{\pm}(y) &= \alpha^{\pm}_1 |{\cal A}|^2 + \alpha^{\pm}_2 \left( \hat{V} + \omega_{xy}^2 \right) + \alpha^{\pm}_3 \hat{V}^2,  \\
\delta \lambda_{\pm}(y) &= \alpha^{s\pm}_1 |{\cal A}|^2 + \alpha^{s\pm}_2 \left( \hat{V} + \omega_{xy}^2 \right) + \alpha^{s\pm}_3 \hat{V}^2. \nonumber
\end{align} 

 The values of the parameters in Eq.~\ref{eqs} are given in table \ref{tb1}. The Hamiltonian in Eq. (\ref{H0}) incorporates the effect of strain on the electronic structure through the parameter ${\cal A}$ and the corresponding gauge field $\textbf{A}_i$. 
 In the effective low-energy Hamiltonian in Eq.(\ref{H0}), an arc-shaped deformation generates an effective gauge field $A_i$ given by $\textbf{A}_i=\eta_i (\hbar/e a_0)\left(\text{Re}[{\cal A}],\text{Im}[{\cal A}]\right)$. The parameters \( \eta_i \) arise from the definition of the pseudogauge fields \( \mathbf{A}_i \), which are expressed in terms of fundamental constants and the strain pattern implemented via \( \mathcal{A} \) \cite{PhysRevB.92.195402}. The strain field is given by \( \mathcal{A} = \varepsilon_{xx} - \varepsilon_{yy} - i 2 \varepsilon_{xy} \), where the in-plane strain tensor components are defined as \( \varepsilon_{ij} = \frac{1}{2} \left( \partial u_i / \partial r_j + \partial u_j / \partial r_i \right) \). Unlike the gauge field associated with a conventional magnetic field which has been studied previously \cite{PhysRevB.99.035416,PhysRevB.100.045411}, the strain-induced gauge field corresponds to opposite magnetic fields in the two valleys and is directly determined by the form of the strain and is not gauge invariant. In addition to the pseudogauge fields, strain induces local shifts in the continuum band structure, given by \( \omega_{xy} = (\partial u_y / \partial x - \partial u_x / \partial y) / 2 \) and \( \hat{V} = \varepsilon_{xx} + \varepsilon_{yy} \). The effects of $\hat{V}$ on excitonic transport have been previously studied\cite{rosati2020strain,dirnberger2021quasi}. 
  
 The effective vector potential resulting from the strain pattern described above is $\textbf{A}_i = \frac{\eta_i \hbar}{ea_0}(y/R, 0)$. This paper focuses on how the emerging gauge field $\textbf{A}_i$ affects excitonic properties, in contrast to the widely studied strain-induced potential $\hat{V}$. 
\begin{table}[h] \label{tb1}
    \begin{tabular}{c c c c c c c c}
\hline\hline 
&eV&&eV&&meV&&meV\\
\hline
$\alpha^{+}_1$&~~15.99&~~~$\alpha^{-}_1$&~~15.92&~~~$\alpha^{s+}_1$ &~~-61&~~~$\alpha^{s-}_1$&~~-5.7\\
$\alpha^{+}_2$&~~ -3.07&~~~$\alpha^{-}_2$&~~-1.36&~~~$\alpha^{s+}_2$&~~3.2&~~~$\alpha^{s-}_2$&~~ 0.02\\
$\alpha^{+}_3$&~~-0.17&~~~$\alpha^{-}_3$&~~ 0.0&~~~$\alpha^{s+}_3$&~~3.4&~~~$\alpha^{s-}_3$&~~ 0.01\\
\hline
\hline
\end{tabular}  
    \caption{Parameters for strain-induced potential terms in Eq. \ref{eqs} \cite{PhysRevB.92.195402}.}
\end{table}
 \begin{figure}[!htbp]
		\centering
		\includegraphics[width=\linewidth]{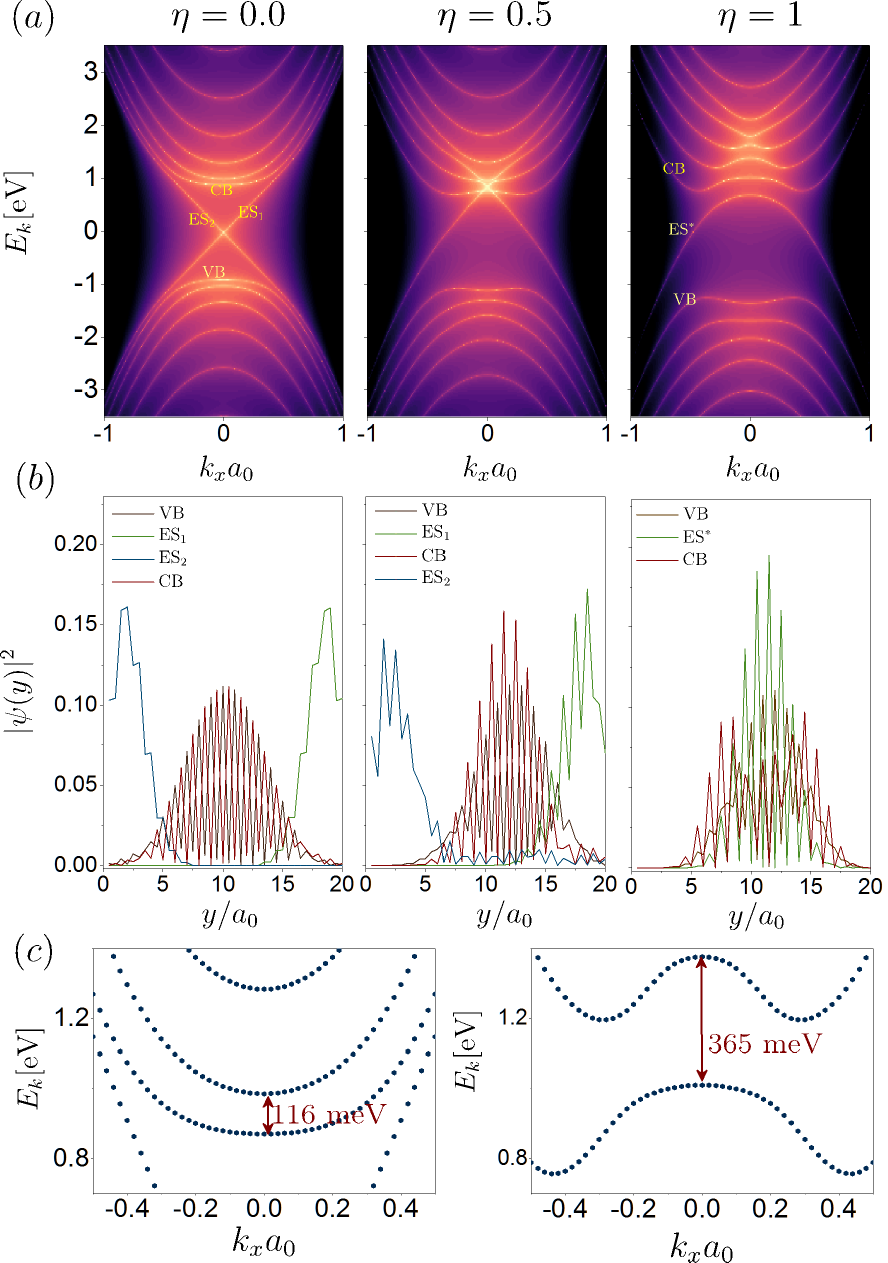} 
        \caption{(a): Quasiparticle energy dispersion of monolayer MoS$_2$ nanoribbon around $k_x = 0$, for  unstrained case, $\eta = 0$, and arc-shaped strained cases, $\eta \neq 0$. We set the width of the nanoribbon to be $L_y=a_0(N+1)$ with $N=20$, and arc radius $R=0.5 L_y$. (b): Probability density $|\psi(y)|^2$ of the conduction (CB) and valence (VB) bands and two edge states, plotted against the coordinate $y \in (0,Na_0)$ with $N=20$ while setting $k_x$ close to zero. At $\eta=1$, the edge states hybridize and extend fully into the bulk, losing their characteristic edge state signature; these hybridized states are denoted as ES$^*$. (c): Comparison of the energy gaps between the first two conduction bands in unstrained ($\eta=0$) and strained systems($\eta=1$): The finite-size gap in the unstrained system is 116 meV ($T=1346$ K), while the strain-induced gap in the strained system increases significantly to 365 meV ($T=4235$ K) for the $\eta=1$ case.}
		\label{fig2}
	\end{figure}
The topological character of the Hamiltonian $H_0$ is determined by the Chern number $2{\cal C} = \sign(\Delta) - \sign(\beta)$\cite{PhysRevB.89.115413}. The non-zero chern number, which is associated with a band inversion, mainly affects the orbital character of the bands rather than general shape of the band structure. In the presence of the strain, the effect of the band inversion (i.e., non-zero Chern number) is more apparent and affects the shape of the band structure by pushing the band minimum away from the corner of the Brillouin zone. To derive the band structure of the electronic states in the strained nanoribbon with respect to the momentum along the nanoribbon, we employ trigonal basis functions $T_n(y) = T(y - y_n)=L_y-|y-y_n|(N+1)$, where $y_n=L_y(\frac{n}{N+1}-\frac{1}{2})$ and $L_y$ is the width of the nanoribbon \cite{PhysRevB.92.195402}. We schematically presented this set of trigonal basis functions $T_n(y)$ in Fig.~\ref{fig1} at the edge of the ribbon in the $y$ direction. This basis is particularly useful since it ensures that the wavefunction vanishes at the nanoribbon boundaries $y = \pm L_y/2$, satisfying the hard-wall boundary conditions imposed by the finite width of the system. These functions also provide sufficient spatial resolution for describing both bulk and edge-localized states, and have been successfully applied in similar continuum models for strained TMDs~\cite{PhysRevB.92.195402}. The number of basis functions $N$ determines the resolution of the numerical method used to solve the Schrodinger differential equation, associated with the continuum Hamiltonian, and is not tied to the number of atoms. We note that quantum confinement effects are incorporated in our method through  hard-wall boundary conditions in the transverse direction, such that the wavefunction satisfies \( \psi_k(y = \pm L_y/2) = 0 \), where \( L_y \) is the width of the nanoribbon. The subband structure shown in Fig. \ref{fig1}(a) is direct result of confinement in the nanoribbon. We consider a nanoribbon width of $N=20$ $(\approx 6.6)$ nm, several times larger than the exciton Bohr radius, while keeping the numerics tractable.  The wave function for an arc-shaped nano-ribbon of TMD is expanded using a set of $T_n(y)$ as
\begin{equation}\label{wf}
    \psi_k^{c,v}(y)=\begin{pmatrix}
        \sum_n a_{n,k}^{c,v} T_n(y) \\
        \sum_n b_{n,k}^{c,v} T_n(y)
    \end{pmatrix},
\end{equation}
 where $v,c$ refer to the valence and conduction bands, respectively. The coefficients $a_{n,k}$ and $b_{n,k}$ represent the spinor components of the wave functions, and $n$ ranges from $0$ to $N$. The orthogonality of the wave function \( \psi_k^{c,v}(y) \) is discussed in the Appendix. The band structure of the strained TMD is derived by diagonalizing the Hamiltonian in the space of the functions in Eq.~(\ref{wf}) which identifies the coefficients $a_{n,k}^{c,v}$ and $b_{n,k}^{c,v}$. 

Fig.~\ref{fig2}(a) shows the energy dispersion of Hamiltonian (\ref{H0}) corresponding to a nanoribbon with zigzag edges under different arc-shaped strain coupling parameters $\eta_i$, where we set $\eta_1 = \eta_2 = \eta_3 = \eta$. As the magnitude of strain increases, the electronic band structure exhibits electron-hole asymmetry and leads to the formation of pseudo-Landau levels \cite{guinea2010energy, levy2010strain, PhysRevLett.113.077201}. The strain-induced pseudogauge fields $A_i$ lead to the localization of the wave function along the $y$ direction, resulting in modification of the energy bands and increased energy spacing between electronic levels \cite{PhysRevB.92.195402}. This localization effect is further illustrated in Fig.~\ref{fig2}(b) shows the probability density $|\psi(y)|^2$ for the conduction (CB) and valence (VB) bands, along with two edge states (ES$_1$ and ES$_2$), corresponding to Fig. \ref{fig2}(a). These are plotted as a function of the coordinate $y \in (0,Na_0)$ along the width, where $N=20$. In the unstrained nanoribbon ($\eta=0$), the edge states are well-defined and localized at the edges. As the strain strength increases, these edge states begin to extend into the bulk. For $\eta=1$, the edge states hybridize and extend fully into the bulk, losing their characteristic edge state signature; we denote these hybridized states as ES$^*$. Strain-induced modifications can extend these edge states into the bulk \cite{guinea2010energy,levy2010strain,VOZMEDIANO2010109,PhysRevB.80.045401,PhysRevB.87.100401}. Given the lack of topological protection due to the opposite parity in the two valleys, we primarily focus on the electronic states in the bulk of the nanoribbon rather than on the edge states in the remainder of the paper. 
The comparison of the energy gaps between the first two bands within the original conduction band in unstrained and strained systems is shown in Fig.~\ref{fig2}.c. In the unstrained system, a finite-size gap of 116 meV arises due to quantum confinement imposed by the hard boundaries of the ribbon. Compared to the room temperature, the finite-size gap may be insufficient to justify the one-dimensional diffusion of excitons \cite{dirnberger2021quasi}. In contrast, the application of strain significantly increases the gap to 365 meV at $\eta = 1$. The strain-induced increase of the gap would support the one-dimensional diffusion of the exciton even at room temperature.

We emphasize that the nanoribbon remains semiconducting despite the appearance of crossing states near zero energy. These are edge-localized modes that exist within the bulk band gap, consistent with the topological properties of zigzag-terminated TMD nanoribbons. Due to the presence of opposite chirality in different valleys, the edge states are not protected. Their apparent protection is due to the fact that we present the low energy spectrum of one individual valley. The bulk states remain gapped across the Brillouin zone. As the ribbon width $L_y$ increases, the edge-state contributions diminish, and the resulting band structure converges to that of a two-dimensional monolayer TMD, which has a finite direct band gap near the $K$ points. This limit confirms that our framework correctly reproduces the known bulk behavior of unstrained 2D TMDs.

%XXXXXXXXXXX

\begin{figure}[!htbp]  
	\centering	\includegraphics[width=\linewidth]{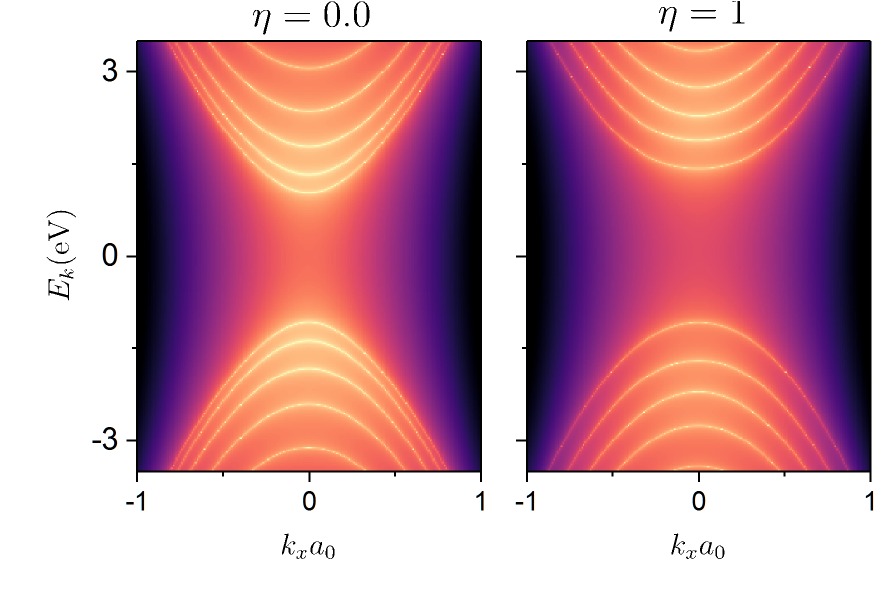} \caption{ Non-topological quasiparticle energy dispersion of monolayer MoS$_2$ nanoribbon around $k_x = 0$, for  unstrained case, $\eta = 0$, and arc-shaped strained cases, $\eta \neq 0$. In this figure, the Chern number is set to zero.} \label{fig8}
\end{figure}

As outlined above, the TMD band structure studied in this paper, harvest topological character in terms of non-zero Chern number $2{\cal C} = \sign(\Delta) - \sign(\beta)$\cite{PhysRevB.89.115413} close to the bottom of the band in each valley. Due to the presence of apposite Chern numbers in the two valleys, the whole band structure is topologically trivial. On the other hand, since circularly polarized light, couples to the valleys individually, the signature of the topological character of individual valley could appear in exciton properties. The topological character of each valley can be tuned by varying the parameters $\alpha$ and $\beta$ of the Hamiltonian and is used to compare the excitonic properties when the electronic Hamiltonian for individual valley is topological trivial.  Fig.~\ref{fig8} demonstrates the non-topological quasiparticle energy dispersion for a monolayer MoS$_2$ nanoribbon under unstrained ($\eta=0$) and strained ($\eta=1$) conditions.  In the non-topological regime, the exciton band structure remains mostly unaffected by the strain, showing only minor modifications such as slight flattening and minor energy shifts. Importantly, no edge states emerge in the non-topological case. This is in stark contrast to the topological case, where strain-induced band inversion and edge-state hybridization result in significant changes to the band structure. In the topological case [Fig.\ref{fig1}], the strain introduces edge states that penetrate the bulk and profoundly modify the system's electronic properties, a feature absent in the non-topological scenario. 

\section{Exciton Band Structure}

In order to derive the exciton band strucutre we need to implement electron-hole interactions in the band structure derived in the last section. The electron-hole interaction leading to the formation of exciton is given by the Hamiltonian \( H_I \):

\begin{equation} \label{H1}
H_{\text{I}} = \int V_{(R-R^\prime)} \left( \sum_{o,o^\prime = a}^b c^\dagger_{R,o} c^\dagger_{R^\prime,o^\prime} c_{R^\prime,o^\prime} c_{R,o} \right) d\mathbf{R} d\mathbf{R^\prime}.
\end{equation}

In the interaction Hamiltonian \( H_I \), \( c^\dagger_{R,o} \) (\( c_{R,o} \)) is the electron creation (annihilation) operator for the spinor \( o \) at the position \( R = (x, y) \). The interaction potential \( V_{(R-R^\prime)} \) describes the Coulomb-like interaction between particles at positions \( R \) and \( R^\prime \).
The reduced Coulomb interaction is described by the Keldysh potential \( V(r) \)~\cite{keldysh,PhysRevB.84.085406} given by:
\begin{align}
V(r)=\frac{\pi e^2}{2 \epsilon r_0} [{\cal H}_0
(r/r_0)-Y_0(r/r_0)].
\end{align}
Here, $\epsilon$ is the dielectric constant of the substrate on which the 2D material lies. For a MoS$_2$ monolayer on a SiO$_2$ substrate, we use $\epsilon=2.5$. The parameter $r_0$ is the screening length given by $r_0=33.875 \text{\AA}/\epsilon$ \ \cite{PhysRevB.89.205436}. ${\mathcal H}_0$ and $Y_0$ denote the Struve function and Bessel function of the second kind of order zero \cite{abramowitz1965handbook}, respectively, and $r$ represents the electron-hole separation distance.

The annihilation operator $c_{r,o}$ is projected into the eigenstates of Hamiltonian (\ref{H0}) as:
 \begin{equation}
     c_{r,o}=\sum_{k_x,E=c,v} e^{-i k_x x} (\sum_n o_n^E T_n(y)) c_{k_x,E},
 \end{equation} 
where $o_n^E$ represents either $a_n^{c,v}$ or $b_n^{c,v}$ components. The Bethe-Salpeter (BS) method \cite{PhysRev.84.1232,PhysRevB.105.195418} given in Eq.~(\ref{BS}) is utilized to derive the exciton dispersion $E_Q$ as a function of the center of mass momentum $Q_x$ along the nanoribbon:

\begin{equation} \label{BS}
    \iint_{-\frac{L_y}{2}}^{\frac{L_y}{2}} dy dy^\prime \sum_{k_x^\prime} \kappa_{vv^\prime}^{cc^\prime}(y,y^\prime,k_x,k_x^\prime,Q) \psi_{k_x^\prime,Q}^{c^\prime,v^\prime}=E_Q \psi_{k_x,Q}^{c,v}.
\end{equation}
In Eq.~(\ref{BS}) the interaction Kernel is given by:
\begin{equation}
\begin{split}
   & \kappa_{vv^\prime}^{cc^\prime}(y-y^\prime,k_x,k_x^\prime,Q)= \\&(E_{Q+k}^c-E_k^v) \delta_{k_xk_x^\prime} \delta_{cc^\prime} \delta_{vv^\prime}- ({\cal D}-{\cal X})_{vv^\prime}^{cc^\prime}(y,y^\prime,k_x,k_x^\prime,Q).
    \end{split}
\end{equation}
Here, $E_k^{c(v)}$ denotes the quasiparticle energy, represented in Fig. \ref{fig1}(a). 
The direct and exchange terms, denoted by ${\cal D}$ and ${\cal X}$, respectively, are given by:
\begin{align}
\mathcal{D}^{cc'}_{fvv'}(y,y',k_x,k_x',Q)
  &= V(y-y',k_x-k_x') \nonumber\\
  &\times W^{v,c',c,v'}_{k_x,\;k_x'+Q,\;k_x,\;k_x'}  % etc.
\end{align}
and 
\begin{equation}
    {\cal X}_{vv^\prime}^{cc^\prime}(y,y^\prime,k_x,k_x^\prime,Q)=  V(y-y^\prime,Q) W^{c^\prime,v,c,v^\prime}_{k_x^\prime+Q,k_x,k_x+Q,k_x^\prime},
\end{equation}
where $V(y-y^\prime,k_x)=\int_0^\infty dx e^{ik_x x} V(y-y^\prime,x)$ is the Fourier transform of the Keldysh potential along the nanoribbon's length. The projection of the Keldysh potential into the electronic structure of the strained nanoribbon is implemented by the functions $W^{E_i}_{q_i}$, which accounts for transitions between valence $(v,v^\prime)$ and conduction $(c,c^\prime)$ bands and is given by:
\begin{equation} 
\begin{split}
    &W^{E_{1,2,3,4}}_{q_{1,2,3,4}}(y,y^\prime)=\\ & \sum_{o,o^\prime=a}^b [(\sum_{n_1} O_{n_1,q_1}^{*} T^*_{n_1}(y)) (\sum_{n_4} O_{n_4,q_4} T_{n_4}(y))]_{E_1E_4}\\ &[(\sum_{n_2} O_{n_2,q_2}^{\prime*} T^*_{n_2}(y^\prime))(\sum_{n_3} O_{n_3,q_3}^{\prime} T_{n_3}(y^\prime))]_{E_2E_3}.
    \end{split}
\end{equation}
Here, $E_i$ and $q_i$ $(i=1,2,3,4)$ represent the energy levels and quantum numbers of the four fermionic operators, respectively. The functions $O_{n,q}$ and $T_n(y)$ are related to the spinor components and the trigonal basis of the wave functions in the transverse direction.
           \begin{figure}[t]
		\centering
		\includegraphics[width=\linewidth]{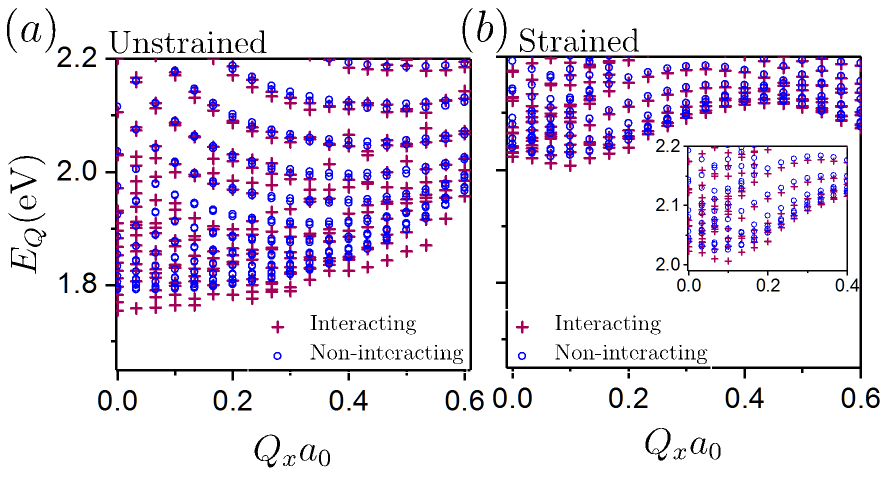} \caption{Exciton energy as a function of center-of-mass momentum $Q_x$
  (scaled by $a_0$), resulting from transitions between the valence band (VB) and conduction band (CB): (a) in the absence of strain and (b) in the presence of strain. The magenta plus symbols represent interacting exciton states, while the blue open circles represent non-interacting exciton states. $N=20$ is used in the calculations. The inset shows a zoomed-in view of the low-energy exciton states.}
		\label{fig3}
	\end{figure}

The exciton dispersion derived using the BS equation (Eq.~\ref{BS}) is presented in Fig.~\ref{fig3}. As discussed in the previous section, strain causes an offset between the conduction band minimum and valence band maximum in momentum space. Fig.~\ref{fig3}(a) illustrates the bulk excitons in the unstrained case ($\eta=0$), while Fig.~\ref{fig3}(b) corresponds to optical transitions between valence and conduction bands under increasing strain. 
When strain is applied [Fig.~\ref{fig3}(b)], the minimum of the exciton dispersion shifts to a finite momentum $Q_x=q_0$ stabilizing momentum-indirect dark excitons whose suppressed radiative recombination naturally leads to longer lifetimes than bright excitons \cite{ind1,ind2,ind3,ind4}. This strain-induced shift of the exciton dispersion minimum results from the combined effect of strain-induced pseudogauge field \cite{PhysRevB.87.165131} and the presence of band inversion in each valley. The interacting exciton states (magenta symbols) exhibit a lower energy than their non-interacting counterparts (blue open circles), highlighting the role of Coulomb interactions in stabilizing excitons and making them more energetically favorable. This energy reduction is evident both in the absence and presence of strain, although strain modifies the overall dispersion characteristics.
 
\begin{figure}[t]
		\centering
\includegraphics[width=\linewidth]{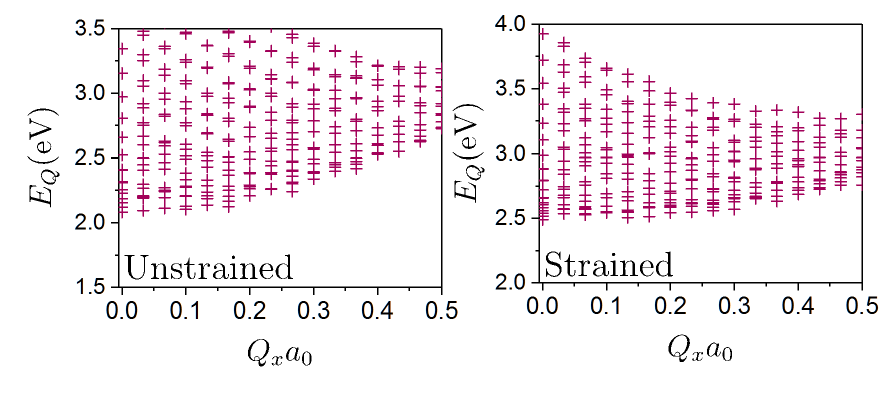} \caption{Exciton energy as a function of center-of-mass momentum $Q_x$ (multiplied by $a_0$) in the non-topological unstrained and non-topological strained cases. In this figure, the Chern number is set to zero.} \label{fig9}
\end{figure}

Fig.~\ref{fig9} compares exciton dispersion for unstrained and strained systems for the non-topological case where $2{\cal C} = \sign(\Delta) - \sign(\beta)=0$. It is evident in this figure that in the absence of band inversion in each valley, the strain has minimal impact on the exciton spectrum, and the energy minimum remains at $Q_x=0$. In contrast, for the topological case, due to the presence of band inversion in each valley, the strain induces a substantial shift of the exciton energy minimum to finite momentum. This feature of the topological case highlights its potential for strain-engineered unidirectional exciton transport, a behavior not observed in the non-topological system.

 \section{Exciton Transport}
\begin{figure}[t]
		\centering		\includegraphics[width=\linewidth]{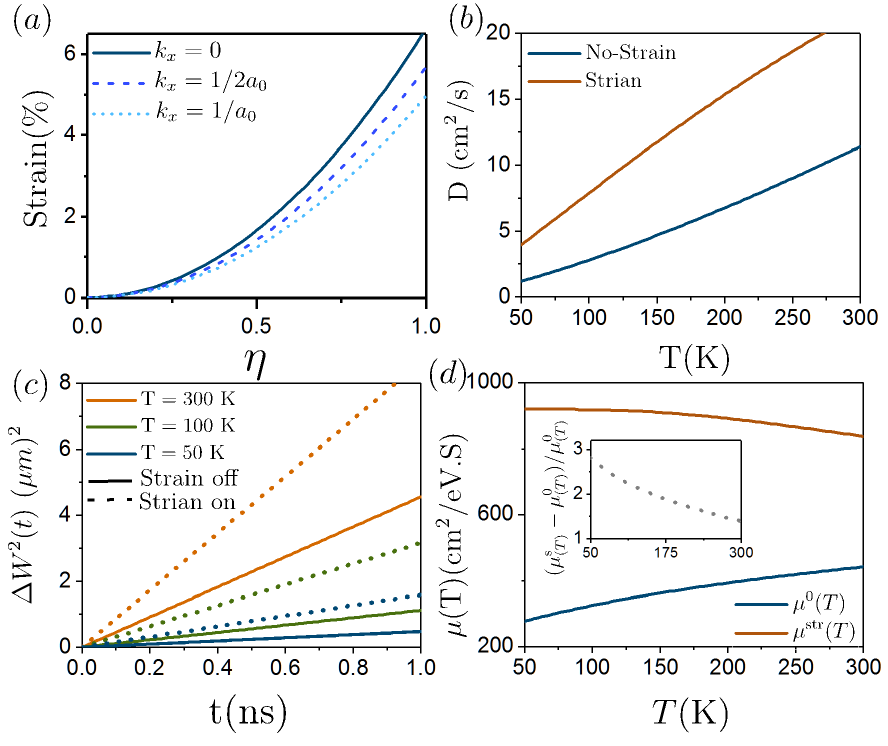} \caption{(a): The strain-induced lattice deformation versus $\eta$. The strain parameter $\eta = 1$ corresponds to a $\sim 6\%$ strain, which is considered a safe strain value for the system without risking structural failure. (b): Temperature dependence of exciton diffusion coefficients under strained and unstrained conditions. The result shows a significant enhancement in exciton diffusion due to applied strain. (c): Time evolution of squared broadening widths of excitons at different temperatures for strained (dashed lines) and unstrained (solid lines) cases. (d): Temperature dependence of exciton mobility for strained (Orange) and unstrained (Blue) cases.  Inset: Relative mobility $(\mu^{\text{str}}(T)-\mu^0(T))/\mu^0(T)$ as a function of temperature, where $\mu^{\text{str}}(T)-\mu^0(T)$ is the mobility difference between strained and unstrained cases.}
		\label{fig4}
	\end{figure}

Strain-induced alterations in the exciton band structure can lead to spatially inhomogeneous exciton diffusion, which is captured by the exciton continuity equation. Two-dimensional TMDs exhibit remarkable mechanical resistance compared to their bulk counterparts, enabling them to sustain significant strain. While typical bulk semiconductors break down under strains below 1\%~, 2D materials can endure deformations exceeding 10\%~\cite{lee2017strain,bertolazzi2011stretching,li2020efficient}. This exceptional mechanical strength allows for substantial manipulation of TMD properties through strain engineering. In our study, we characterize the mechanical strain in TMD nanoribbons using the tuning parameter $\eta$ in the model Hamiltonian. At $\eta = 1$, the strain reaches 6\%~ [Fig.~\ref{fig4}(a)], a value well within the typical strain tolerance of TMDs before structural failure. This strain results in a shift of the minimum of the exciton dispersion to a finite center-of-mass momentum, as shown in Fig.~\ref{fig3}(b).
Strain-induced alterations to the exciton band structure leads to spatially inhomogeneous exciton distributions that can be described by the exciton continuity equation:
 \cite{rosati2020strain}
\begin{equation} \label{Exc_Eq}
\frac{\partial n(x, t)}{\partial t} = \tau \sum_Q \frac{\partial}{\partial x} \left( v_Q \frac{\partial}{\partial x} \left[ v_Q \, n(x, t) \, \frac{e^{-E_Q/k_B T}}{Z} \right] \right),
\end{equation}

Here, \( n(x, t) \) denotes the exciton density as a function of position along the nanoribbon and time, and \( v_Q = \frac{1}{\hbar} \frac{\partial E_Q}{\partial Q} \) is the group velocity of excitons with center-of-mass momentum \( Q \). Since the system is translationally invariant along the \( x \)-direction and finite along the \( y \)-direction, exciton dynamics is effectively one-dimensional. The exciton distribution is assumed to be Gaussian along the transverse \( y \)-direction, while all transport occurs along the longitudinal axis \( x \). The partition function is given by \( Z = \sum_Q e^{-E_Q/k_B T} \). We emphasize that this 1D form of the continuity equation aligns with the nanoribbon geometry considered in our model.
 
      \begin{figure}[t]
		\centering		\includegraphics[width=\linewidth]{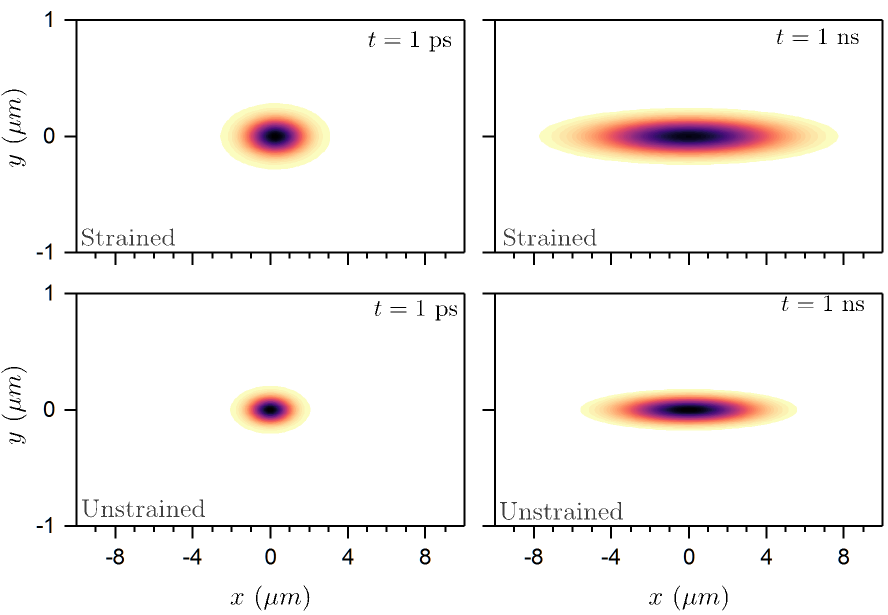} \caption{Spatio-temporal evolution of exciton populations $n(r,t)$ in nanoribbon along x at room temperature $T=300$ K. The comparison between strained (top) and unstrained (bottom) conditions demonstrates enhanced exciton diffusio induced by strain, measured at times $t=1$ ps and $t=1$ ns. The diffusion in the strained (unstrained) case is caused by transitions from ES$^*$ (ES$_1$) to CB.}
		\label{fig5}
	\end{figure}
    
    Using the modified Fick's second law for excitons, i.e., $ \partial_t n(r,t)={\cal D} \nabla^2 n(r,t)$ \cite{rosati2020strain,thompson2022anisotropic}, the diffusive motion is described by the diffusion coefficient $D$ given by:
 \begin{equation}
    {\cal D}=\sum_{Q} \tau v^2_{Q} \dfrac{e^{-E_Q/k_B T}}{{\cal Z}}.
 \end{equation}
 Fig.~\ref{fig4}(b) illustrates the temperature dependence of the diffusion coefficient enhancement along the nanoribbon after strain is applied. The effect of strain manifests as modifications to the group velocity and exciton energy spectrum $E_Q$, resulting from strain-induced band structure modulations.

The spatio-temporal evolution of exciton populations is described by the solution of Eq.~\ref{Exc_Eq}. For an initial delta function distribution $n(r)=n_0 \delta(r)$, the solution is given by a Gaussian function, $n(r,t)=(n_0/\sqrt{\pi \Delta w^2(t)}) e^{-r^2/\Delta w^2(t)}$, where $\Delta w^2(t)=w^2(t)-w_0^2=4Dt$ is the squared broadening width. Fig. \ref{fig4}(c) illustrates the time evolution of the squared broadening width $w^2(t)$ of excitons along the nanoribbon. The squared broadening width quantifies the spatial spread of the exciton population over time. The linear relationship between $\Delta w^2(t)$ and $t$ allows for the experimental determination of the diffusion coefficient by measuring the time-dependent spread of the exciton population \cite{dirnberger2021quasi,PhysRevLett.120.207401,he2015spatiotemporal,rosati2020strain}.  
 
The application of arc-shaped strain to a TMD nanoribbon can also affect carrier mobility. In the unstrained case, excitons typically occupy the lowest energy state at zero momentum ($q_0$ = 0), with mobility primarily determined by diffusion: $\mu_0(T) = D_0(T)/k_B T$ 
\cite{feynman2015feynman}, where $D_0$ is the unstrained diffusion coefficient. However, the shift of the exciton energy minimum to a finite momentum $s q_0$ where $s=\pm 1$, leads to an extra mobility term given by \ $\mu_d^s(q_0) = v_{d}^s(q_0)/|\nabla_y E_Q|$, which leads to further spread of the excitons. Note that since the extra diffiusion vanishes after summing over $s$ it does not lead to net transprot of excitons as expected. The strain-induced mobility would be $\mu(T)=\sum_s D^{s}(T)/k_BT,$
where $D^{s}(T)$ is the diffusion coefficient in the presence of strain.

Fig.~\ref{fig4}(d) presents the temperature-dependent exciton mobility in both strained and unstrained nanoribbons. Our results reveal consistently higher mobility values in the strained case across all temperatures. The inset quantifies this enhancement by showing the relative mobility increase, defined as $(\mu^{\text{s}}(T)-\mu^0(T))/\mu^0(T)$, demonstrating a significant strain-induced mobility enhancement that varies with temperature.
 
Fig.~\ref{fig5} shows the spatio-temporal evolution of exciton populations $n(r,t)$ along nanoribbon at room temperature, with and without strain. The results demonstrate significant enhancement in exciton diffusion within strained nanoribbons. Due to the finite size gap between the pseudo Laundau levels, we modeled the exciton distribution with a time-independent Gaussian profile along the y-axis.  The temporal evolution of exciton distribution along x-axis is modeled by the solution of Eq.~\ref{Exc_Eq}. The enhanced diffusion observed under strain, driven by the strain-induced modifications to the band structure, contrasts with the slower exciton diffusion observed in the unstrained case. 

\begin{figure}[t]
	\centering
\includegraphics[width=\linewidth]{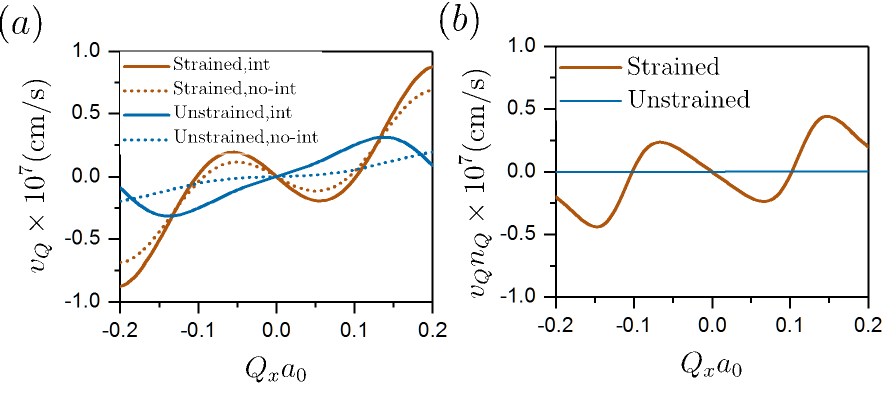} \caption{(a) Group velocity of excitons ($v_Q$) for strained (brown) and unstrained (blue) systems when interactions between excitons is on (solid line) and off (dotted line) extracted from Fig.\ref{fig3}. Solid lines represent the presence of interactions, while dashed lines indicate their absence. (b) Product of the bosonic distribution function ($n_Q$) and group velocity ($v_Q$) as a function of the wavevector ($Q_x a_0$), showing significant oscillations under strain (brown) and negligible values for the unstrained case (blue).} \label{fig6}
\end{figure}
The underlying mechanism for the impact of strain on exciton transport is examined by analyzing the product of the bosonic distribution function $n_Q$ and the group velocity of excitons $v_Q=(1/\hbar)(\partial_{Q}E_Q)$. Our results are presented in Fig.~\ref{fig6}(b), which demonstrates that in the unstrained system (blue curves), the product $n_Q v_Q$ remains nearly constant and close to zero across all wavevectors. Under the strain (brown curves), significant variation in $n_Q v_Q$ emerge, suggesting the formation of local exciton currents. These variations exhibit an alternating pattern of positive and negative values of $n_Q v_Q$ around $Q = 0$, leading to the strain-induced modulation of exciton transport.  Furthermore, Fig.~\ref{fig6}(a) reveals the role of strong electron-hole interactions in shaping exciton dynamics. The solid and dotted lines correspond to cases with and without Coulomb interactions, respectively. In the strained system, interactions enhance the magnitude of the exciton group velocity $v_Q$, leading to stronger unidirectional exciton diffusion. 

\section{Exciton dipole moment}

Strong exciton-exciton interactions are crucial for the realization of novel excitonic phases and potential applications. As outlined in the introduction, the shape of the electronic wave functions and their dependence on momentum along the nanoribbon can lead to a larger in-plane dipole moment, which in turn enhances exciton-exciton interactions. This effect can be quantified by estimating the exciton dipole moment, \( d_{\text{ex}} \) expressed as:
\begin{equation} \label{dipol}
d_{\text{ex}} = -e \int y \left[ |\psi_e(y)|^2 \Big|_{k_e^{\min}} - |\psi_h(y)|^2 \Big|_{k_h^{\max}} \right] dy %= -e \langle r_e - r_h \rangle.
\end{equation}
 The terms \( |\psi_e(y)|^2 \) and \( |\psi_h(y)|^2 \) represent the probability distributions of the electron and hole, respectively, with their respective wave vectors along the nano-ribbon being at \( k_e^{\min} \) and \( k_h^{\max} \), which correspond to the minimum (for the electron) and maximum (for the hole) of the conduction (valence) band. The integral quantifies the net displacement between the electron and hole, determining the magnitude of the dipole moment. Since the exciton dispersion minimum in the presence of strain occurs at a finite center-of-mass momentum, the electron and hole components acquire different momenta. Due to the particular structure of states under a strain-induced gauge field, an electron and hole with different momenta along the nanoribbon correspond to different spatial profiles of their wavefunctions along its width. Such a structure results in a finite in-plane dipole moment. Numerical calculations based on Eq.~\ref{dipol} indicate that strain significantly enhances the dipole moment, increasing it from $d_{ex}\approx 0.5 a$
 in the unstrained case to $d_{ex}\approx 8 a_0$ under strain, where $a_0$ is the MoS$_2$ lattice parameter. This enhancement suggests that strain modifies carrier confinement, leading to greater electron-hole separation. We should note that the dependence of exciton dipole moments on their center of mass momentum, is a unique feature of excitons in strained TMDs. 
 \begin{figure}[t] 
		\centering
\includegraphics[width=\linewidth]{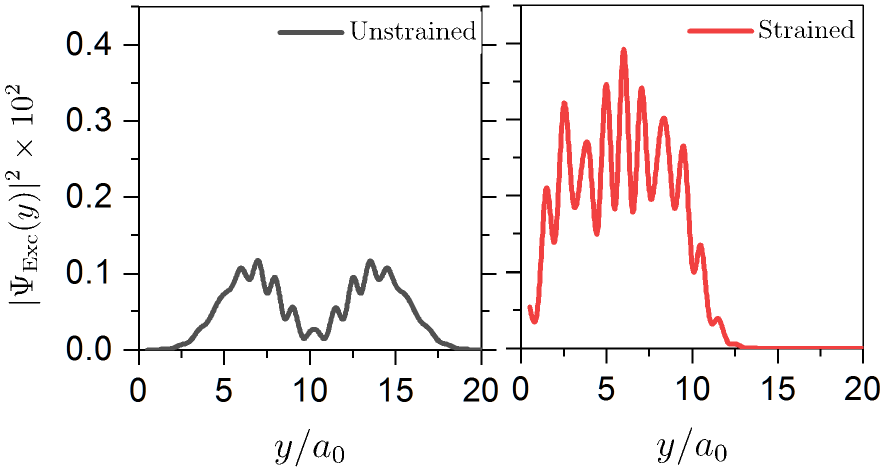} \caption{\label{fig7}Exciton probability density $|\Psi_{\mathrm{Exc}}(y)|^2$ across the transverse direction ($y/a_0$) of the nanoribbon for strained (right panel, $\eta = 1$) and unstrained ( left panel, $\eta = 0$) cases. The figure indicates the spatial distribution of the exciton wavefunction, revealing enhanced asymmetry and localization under strain.}
\end{figure} 

 To clarify the spatial nature of the exciton dipole moment discussed above, we compute the exciton wavefunction $\psi(\mathbf{r}_e, \mathbf{r}_h)$ using the Bethe–Salpeter formalism. It is constructed as a coherent superposition of electron–hole pair states:
\begin{equation}
\psi_{\text{Exc}}(\mathbf{r}_e, \mathbf{r}_h) = \sum_{vc} A_{vc}\, \phi_v^*(\mathbf{r}_h)\, \phi_c(\mathbf{r}_e),
\end{equation}
where $A_{vc}$ are the eigenvector components from the Bethe–Salpeter equation, and $\phi_{v}$ and $\phi_{c}$ are the single-particle valence and conduction band wavefunctions, respectively. To visualize the exciton spatial distribution across the ribbon, we project the two-particle wavefunction onto the transverse direction as
\begin{equation} \label{WExc}
|\Psi_{\text{Exc}}(y)|^2 = \sum_{x_e, x_h} |\psi_{\text{Exc}}(x_e, y; x_h, y)|^2.
\end{equation} 
Eq. \ref{WExc} provides the projected exciton probability distribution, from which the electron–hole separation in Eq. 15 is inferred.
Fig~\ref{fig7} shows the resulting exciton probability density along \( y/a_0 \), demonstrating stronger localization under strain.
 According to Fig. \ref{fig7} for the unstrained system ($\eta = 0$), the wavefunction is symmetric and delocalized, indicating strong electron–hole overlap and a negligible dipole moment. Under strain ($\eta = 1$), however, the wavefunction becomes asymmetric and localized near one edge of the nanoribbon, revealing strain-induced spatial separation between the electron and hole. This behavior arises from the strain-modified band structure and pseudogauge field effects, and explains the large in-plane dipole moment computed in the strained regime.

 \section{Conclusion}
We explored the impact of strain on exciton dynamics in TMD nanoribbons. The effects of strain on the low-energy electronic band structure of TMDs manifest as local potentials (which shift the band structure for all momenta) and a pseudo-gauge field (which modifies the dispersion relation). Using the Bethe-Salpeter (BS) formalism, we derived the exciton dispersion in the presence of arc-shaped strain.
By increasing the energy gap between states within the valence or conduction bands relative to the finite-size gap and shifting the exciton dispersion minimum to a finite center-of-mass momentum, strain extends the exciton lifetime and enhances unidirectional diffusion. Furthermore, excitons in strained TMDs develop a sizable in-plane dipole moment, leading to strong exciton-exciton interactions.
Our results demonstrate that strain engineering can be leveraged to design excitonic devices and circuits, where generated excitons diffuse while remaining dark due to their finite center-of-mass momentum and subsequently decay in unstrained regions of the sample.

 The formulation of the Bethe-Salpeter equation for excitonic states in a trigonal over-complete basis enables a comprehensive analysis of exciton binding energies and wavefunctions within strain-modulated band structures, where translational symmetry is broken in one direction.
Our results on exciton diffusion in strained samples demonstrate how band structure modifications lead to enhanced unidirectional diffusion through increased mobility. Furthermore, our findings reveal that the topological characteristics of each TMD valley play a crucial role in excitonic transport and can potentially be identified through the study of excitonic band structures.
Additionally, our method can be applied to different strain patterns to design excitonic circuits. 

Given the particular in-plane anisotropic dipole moment of excitons, which also depends on the center-of-mass momentum, strained samples could stabilize novel interacting phases. The study of such interacting excitonic phases will be the focus of future research.

\section{Data Availability Statement }
The data that support the findings of this study are available from the authors upon reasonable request.

 \section{ACKNOWLEGEMENTS}
 We acknowledge support from NSF grants DMR 2315063 (PG) and DMR 2130544 (PG and SH). This research was supported in part by grant NSF PHY-2309135 to the Kavli Institute for Theoretical Physics (KITP). We are thankful for fruitful discussions with M. Hafezi, A. Strivitsava, A. Grankin and R. Asgari.
\appendix*
\section{Orthonormality in a Non-Orthogonal Basis} 
For a TMD nanoribbon with zigzag edges, the wave function can be expanded using a set of trigonal functions $T_n(y)$ as
\begin{equation}
    \psi_E(y)=\begin{pmatrix}
        \sum_n a_n^E T_n(y)\\
        \sum_n b_n^E T_n(y)
    \end{pmatrix}, 
\end{equation}
where $E = v,c$ denotes the valence and conduction bands, respectively. The index $n$ ranges from $0$ to $N$, with $N$ representing the number of basis functions. The coefficients \( a_n^E \) and \( b_n^E \) are obtained from solving the generalized quasi one-dimension Bethe–Salpeter equation (BSE) in Eq. \ref{BS}. The continuum Hamiltonian in Eq.~\ref{H0} is projected onto the non-orthogonal trigonal basis functions \( T_n(y) \), leading to a generalized Hermitian eigenvalue problem of the form \( H \psi_E(y) = E \psi_E(y) \). Solving this eigenvalue problem yields the energy spectrum and the corresponding expansion coefficients used throughout the paper.

It is important to note that these trigonal basis functions do not satisfy the orthogonality condition. Their overlap is given by \cite{PhysRevB.92.195402}
\begin{equation}
\braket{T_m}{T_n} = \frac{L_y^3}{N+1} \left[\frac{2}{3} \delta_{m,n} + \frac{1}{6}(\delta_{m,n+1} + \delta_{m,n-1})\right]
\end{equation}
where $L_y=a_0 (N+1)$ is the width of the nanoribbon, $\delta_{i,j}$ is the Kronecker delta function, and the overlap integral is defined as $\braket{T_m}{T_n}=\int dy \ T_m^*(y) T_n(y)$.

The inner product between two states is then:
\begin{equation} \label{inner}
\langle \psi_E | \psi_{E'} \rangle = \sum_{n,m} \left( a_n^{E*} a_m^{E'} + b_n^{E*} b_m^{E'} \right) \braket{T_m}{T_n}.
\end{equation}
Since the wave function \( \psi_E(y) \) is orthonormal, i.e., $\langle \psi_E | \psi_{E'} \rangle=\delta_{E,E^\prime}$, and satisfy the eigenvalue problem, then the expansion coefficients in Eq.~\ref{inner} satisfies:
\begin{equation} \label{ab}
\left(a^{E*}_n a^{E'}_m + b^{E*}_n b^{E'}_m\right) \propto \ \dfrac{N+1}{L_y^3} (\dfrac{3}{2} \delta_{n,m}+6 \delta_{n,m\pm 1}) \delta_{E,E^\prime}
\end{equation}
which clearly represents orthogonality when $E \neq E^\prime$. This relation has also been verified numerically by computing overlaps between numerically obtained eigenstates.
To verify Eq.~\ref{ab} numerically, we compute the inner product \( \langle \psi_E | \psi_{E'} \rangle \) using the expansion coefficients \( \{a_n^E, b_n^E\} \) obtained from solving the generalized eigenvalue problem. The resulting overlaps match \( \delta_{E,E'} \) to machine precision, confirming the orthonormality of the eigenstates.
\\
In conclusion, even though the trigonal basis functions \( T_n(y) \) are not orthogonal, the wavefunctions \( \psi_E(y) \) constructed from them form an orthonormal set. This is because the expansion coefficients \( \{a_n^E, b_n^E\} \) are solutions to a generalized Hermitian eigenvalue problem that incorporates the nontrivial overlap matrix.

\bibliography{ref}
\end{document}